\documentclass[a4paper]{llncs}
\usepackage{slashbox}
\usepackage{amssymb}
\usepackage{graphicx}
\usepackage{array}
\usepackage{xspace}
\usepackage{url}
\usepackage{color}
\usepackage{multirow}

\usepackage{amsmath}
\usepackage{color}
\usepackage{amssymb}
\usepackage{algorithmic}
\usepackage[ruled,linesnumbered,algo2e]{algorithm2e}
\usepackage{array}
\usepackage{booktabs}
\usepackage{color}
\usepackage{colortbl}
\usepackage{epsfig}
\usepackage{fancybox}
\usepackage{graphicx}
\usepackage{helvet}
\usepackage{listings}
\usepackage{multirow}
\usepackage{pifont}
\usepackage{rotate}
\usepackage{amssymb,amsmath}
\usepackage{url}
\usepackage{listings}
\usepackage{paralist}

\usepackage{subfigure}
\usepackage{tabto}

\newcommand{\specialcell}[2][t]{\begin{tabular}[#1]{@{}l@{}}#2\end{tabular}}
\newcommand{\framework}{\textit{Predictive Business Process Monitoring Framework}\xspace}

\urldef{\mailsc}\path||
\newcommand{\keywords}[1]{\par\addvspace\baselineskip
\noindent\keywordname\enspace\ignorespaces#1}

\begin{document}

\title{Predictive Monitoring of Business Processes}

\author{Fabrizio Maria Maggi\inst{1} \and Chiara Di Francescomarino\inst{2} \and \\ Marlon Dumas\inst{1} \and Chiara Ghidini\inst{2}}
\institute{University of Tartu, Liivi 2, 50409 Tartu, Estonia.\\
\email{\{f.m.maggi, marlon.dumas\}@ut.ee}
\and
FBK-IRST, Via Sommarive 18, 38050 Trento, Italy.\\
\email{\{dfmchiara, ghidini\}@fbk.eu}}
\maketitle

\begin{abstract}
Modern information systems that support complex business processes generally maintain significant amounts of process execution data, particularly records of events corresponding to the execution of activities (event logs).
In this paper, we present an approach to analyze such event logs in order to predictively monitor business goals during business process execution.
At any point during an execution of a process, the user can define business goals in the form of linear temporal logic rules.
When an activity is being executed, the framework 
identifies input data values that are more (or less) likely to lead to the achievement of each business goal.
Unlike reactive compliance monitoring approaches that detect violations only after they have occurred, our predictive monitoring approach provides early advice so that users can steer ongoing process executions towards the achievement of business goals. In other words, violations are predicted (and potentially prevented) rather than merely detected.
The approach has been implemented in the ProM process mining toolset and validated on a real-life log pertaining to the treatment of cancer patients in a large hospital.

\end {abstract}

\section{Introduction}
\label{sec:intro}

The execution of business processes is generally subject to internal policies, norms, best practices, regulations, and laws.
For example, a doctor may only perform a certain type of surgery if this is preceded by a pre-operational screening,
while in a sales process, an order can be archived only after that the customer has confirmed receipt of all ordered items.
We use the term \emph{business constraint} to refer a requirement imposed on the execution of a process that separates compliant from non-compliant behavior \cite{Pesic2006:ConDec}.



Compliance monitoring is an everyday imperative in many organizations. Accordingly, a range of research proposals have addressed the problem of monitoring business processes with respect to business constraints \cite{Maggi2011,MaggiRV,DBLP:conf/fase/MaggiMA12,santos,Ly11,DBLP:journals/tweb/MontaliPACMS10,benatallah,Holmes11,Birukou10,Weidlich11}. Given a process model and a set of constraints -- expressed, e.g., in temporal logic -- these techniques provide a basis to monitor ongoing executions of a process (a.k.a.\ \emph{cases}) in order to assess whether they comply with the constraints in question. 
However, these monitoring approaches are \emph{reactive}, in that they allow users to identify a
violation only \emph{after it has occurred} rather than supporting them in \emph{preventing} such violations in the first place.


In this setting, this paper presents a novel monitoring framework, namely \emph{Predictive Business Process Monitoring}, based on the continuous generation of predictions and recommendations on what activities to perform and what input data values to provide, so that the likelihood of violation of business constraints is minimized. 
At any point during the execution of a business process, the user can specify a \emph{business goal} using Linear Temporal Logic (LTL).\footnote{In line with the forward-looking nature of predictive monitoring, we use the term \emph{business goal} rather than \emph{business constraint} to refer to the monitored properties.} Based on an analysis of execution traces, the framework continuously provides the user with estimations of the likelihood of achieving each business goal for a given ongoing process execution. The proposed framework takes into account the fact that predictions
often depend both on: (i) the sequence of activities executed in a given case; and (ii) the values of data attributes after each activity execution in a case. For example, for some diseases, doctors may decide whether to perform a surgery or not, based on the age of the patient, while in a sales process, a discount may be applied only for premium customers.



The core of the proposed framework is a method to generate predictions of business goal fulfillment. Specifically, the technique estimates for each enabled activity in an ongoing case, and for every data input that can be given to this activity, the probability that the execution of the activity with the corresponding data input will lead to the fulfillment of the business goal. In line with the principle of considering both control-flow and data, the proposed technique proceeds according to a two-phased approach. Given an ongoing case in which certain activities are enabled, we first select from the set of completed execution traces, those that have a prefix ``similar'' to the (uncompleted) trace of the ongoing case (control-flow matching). Next, for each selected trace, we produce a \emph{data snapshot} consisting of a value assignment for each data attribute up to its matched prefix. Given a business goal, we classify a data snapshot as a positive or a negative example based on whether the goal was eventually fulfilled in the completed trace or not. In this way, we map the prediction task to a classification task, wherein the goal is to determine if a given data snapshot leads to a business goal fulfillment and with what probability. Finally, we solve the resulting classification task using decision tree learning, i.e., we produce a decision tree to discriminate between fulfillments and violations. The decision tree is then used to estimate the probability that the business goal will be achieved, for each possible combination of input attribute values.

The proposed framework can be applied both for prediction and recommendation. For prediction, the decision tree is used to evaluate the probability for the business goal to be achieved for a given combination of attribute values. For recommendation, the decision tree is used to select combinations of attribute values that maximize the probability of the business goal being achieved.

The predictive monitoring framework has been implemented in the ProM toolset for process mining. The framework has been validated using a real-life log (provided for the 2011 BPI challenge~\cite{bpichallenge2011}) pertaining to the treatment of cancer patients in a large Dutch academic hospital.

The remainder of the paper is structured as follows. Section~\ref{sec:motivation} introduces a running example.
Section~\ref{sec:background} introduces concepts pertaining to LTL and decision trees.
Section~\ref{sec:approach} presents the predictive monitoring framework and its implementation.
Section~\ref{sec:evaluation} discusses the validation on a real-life log.
Finally, Section~\ref{sec:related} discusses related work and Section~\ref{sec:conclusion} draws conclusions and perspectives. 
\section{Running Example}
\label{sec:motivation}

During the execution of a business process, process participants cooperate to achieve certain business goals. At any stage of the process enactment, decisions are taken aimed at reaching these goals. Therefore, it becomes crucial for process participants to be provided with predictions on whether the business goals will be achieved or not and, even more, to receive recommendations about the choices that maximize the probability of reaching the business goals.

\begin{figure}
	\begin{center}
		\includegraphics[width=\textwidth]{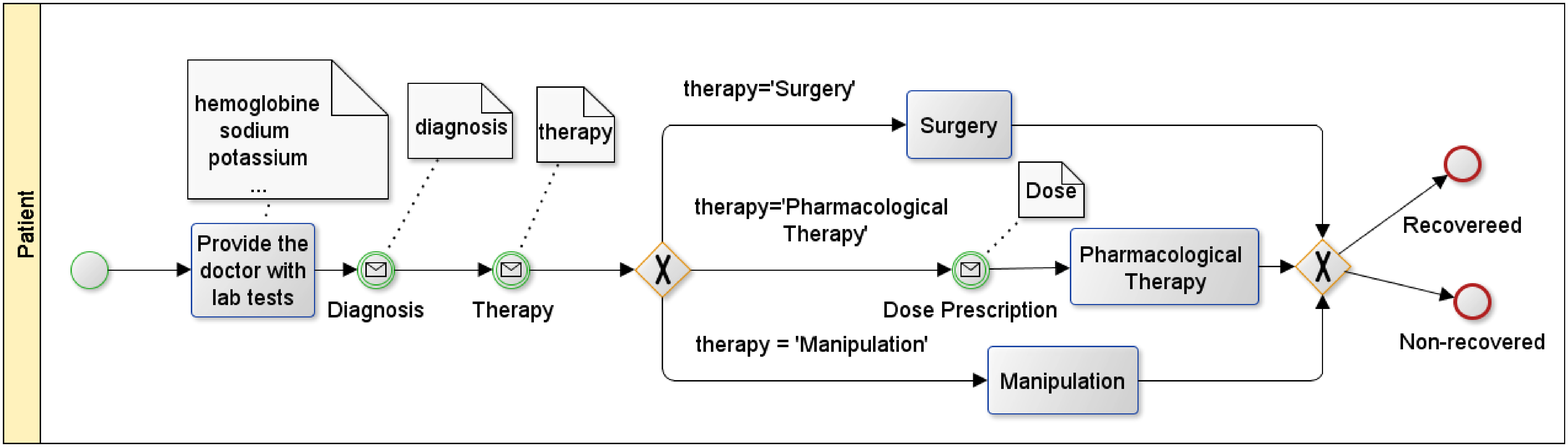}
	\end{center}
	\caption{A simple process describing the medical treatment management.}
	\label{fig:run_ex}
\end{figure}

\figurename~\ref{fig:run_ex} shows a BPMN model of a business process we will use as running example. It describes how a patient is nursed according to the instructions of a doctor. During the process execution, the doctor has to make decisions on therapies and on the doses of medicines to be administered to the patient.
The process starts when the patient provides the doctor with lab test results. Based on the tests, the doctor formulates a diagnosis. Then, the doctor has to decide the therapy to prescribe. The therapy can be a surgery, a pharmacological therapy or a manipulation. In case of a pharmacological therapy, the doctor has also to prescribe the quantity of medicine the patient has to assume.

In this scenario, historical information about past executions of the process could be used to support the doctor in making decisions by providing him or her with predictions about the (most likely) iter of the disease and recommendations about the best choices to be made in order to guarantee the patient recovery. The approach presented in this paper aims at supporting process participants in their decisions by providing them with predictions about the realization of their goals and, in case they can influence the process with their decisions, by recommending them the best choices to be made to achieve their business goals.

In our example, the goal of the doctor could be that every diagnosis is eventually followed by the patient recovery. By exploiting data related to the clinical history of other patients with similar characteristics, our technique aims at providing the process participants with predictions about whether the patient will recover or not. In addition, whenever the doctor has to make decisions (e.g., prescribe the type of therapy or choose the dose of a medicine), recommendations are provided about the options for which it is more likely that the patient will recover.


\section{Background}
\label{sec:background}
\newcommand{\lnext}{\ensuremath{\mathbf{X}}}
\newcommand{\lwnext}{\ensuremath{\mathbf{\bar{X}}}}
\newcommand{\luntil}{\ensuremath{\mathbf{U}}}
\newcommand{\lsince}{\ensuremath{\mathbf{S}}}
\newcommand{\lzero}{\ensuremath{\mathbf{Z}}}
\newcommand{\ltrigger}{\ensuremath{\mathbf{T}}}
\newcommand{\lrelease}{\ensuremath{\mathbf{R}}}
\newcommand{\lwuntil}{\ensuremath{\mathbf{W}}}
\newcommand{\lglobally}{\ensuremath{\mathbf{G}}}
\newcommand{\lfuture}{\ensuremath{\mathbf{F}}}
\newcommand{\tnext}{\ensuremath{\mathbf{X}_{[t_1,t_2]}}}
\newcommand{\twnext}{\ensuremath{\mathbf{\bar{X}_I}}}
\newcommand{\tuntil}{\ensuremath{\mathbf{U}_{[t_1,t_2]}}}
\newcommand{\tsince}{\ensuremath{\mathbf{S}_{[t_1,t_2]}}}
\newcommand{\trelease}{\ensuremath{\mathbf{R}_{[t_1,t_2]}}}
\newcommand{\tglobally}{\ensuremath{\mathbf{G}_{[t_1,t_2]}}}
\newcommand{\lonce}{\ensuremath{\mathbf{O}}}
\newcommand{\tonce}{\ensuremath{\mathbf{O}_{[t_1,t_2]}}}
\newcommand{\lyesterday}{\ensuremath{\mathbf{Y}}}
\newcommand{\tyesterday}{\ensuremath{\mathbf{Y}_{[t_1,t_2]}}}
\newcommand{\lhistorically}{\ensuremath{\mathbf{H}}}
\newcommand{\thistorically}{\ensuremath{\mathbf{H}_{[t_1,t_2]}}}
\newcommand{\tfuture}{\ensuremath{\mathbf{F}_{[t_1,t_2]}}}
\newcommand{\nfuture}{\ensuremath{\mathbf{F}_{[0,t_1]}}}
\newcommand{\true}{\ensuremath{\mbox{true}}}
\newcommand{\false}{\ensuremath{\mbox{false}}}
\newcommand{\n}{\ensuremath{\figitem{N}}}
\newcommand{\old}{\ensuremath{\figitem{O}}}
\newcommand{\R}{\ensuremath{\mathbf{R}_+}}
In this section, we first introduce the language used for business goal definition (LTL), followed by an an overview on decision tree learning.

\subsection{LTL}
\label{sec:ltl}
In our proposed approach, a business goal can be formulated in terms of LTL rules, as LTL (and its variations) is classically used in the literature for expressing business constraints on procedural knowledge~\cite{Pesic2007:DECLARE}. LTL~\cite{Pnueli:1977} is a modal logic with modalities devoted to describe time aspects.
Classically, LTL is defined for infinite traces. However, when focusing on the compliance of business processes, we use a variant of LTL defined for finite traces (since business process are supposed to complete eventually).

We assume that events occurring during the process execution fall in the set of atomic propositions. LTL rules are constructed from these atoms by applying the temporal operators $\lnext$ (next), $\lfuture$ (future), $\lglobally$ (globally), and $\luntil$ (until) in addition to the usual boolean connectives. Given a formula $\varphi$, $\lnext \varphi$ means that the next time instant exists and $\varphi$ is true in the next time instant (strong next). $\lfuture \varphi$ indicates that $\varphi$ is true sometimes in the future. $\lglobally \varphi$ means that $\varphi$ is true always in the future. $\varphi \luntil \psi$ indicates that $\varphi$ has to hold at least until $\psi$  holds and $\psi$ must hold in the current or in a future time instant.

In the context of the running example, examples of relevant business goals formulated in terms of LTL rules include:
\begin{itemize}
\footnotesize
 	\item $\varphi_0 = 	\lglobally (``diagnosis" \rightarrow \lfuture(``recovered"))$,
  \item $\varphi_1 = \lfuture(``tumor~marker~CA-19.9") \vee  \lfuture(``ca-125~using~meia")$,
  \item $\varphi_2 = \lglobally(``CEA-tumor~marker~using~meia" \rightarrow$  \\ \tabto{1.5cm} $\lfuture(``squamous~cell~carcinoma~using~eia"))$,
  \item $\varphi_3 = (\neg``histological~examination-biopsies~nno")$ \\ \tabto{1.5cm} $\luntil (``cytology-ectocervix-")$,
  \item $\varphi_4 = \lfuture(``histological~examination-big~resectiep")$, and
  \item $\varphi_5 = (\neg``histological~examination-biopsies~nno")$ \\ \tabto{1.5cm} $\luntil (``squamous~cell~carcinoma~using~eia")$.
\end{itemize}


\subsection{Decision Tree Learning}
\label{sec:decision}

%

Decision tree learning uses a decision tree as a model to predict the value of a target variable based on input variables (features). Decision trees are built from a set of training dataset.
Each internal node of the tree is labeled with an input feature. Arcs stemming from a node labeled with a feature are labeled with possible values or value ranges of the feature.
Each leaf of the decision tree is labeled with a class, i.e., a value of the target variable given the values of the input variables represented by the path from the root to the leaf.

Each leaf of the decision tree is associated with a class support (\textit{class support}) and a probability distribution (\textit{class probability}). \textit{Class support} represents the number of examples in the training set, that follow the path from the root to the leaf and that are correctly classified; \textit{class probability} ($prob$) is the percentage of examples correctly classified with respect to all the examples following that specific path, as shown in the formula reported in (\ref{eq:confidence}).
\begin{equation}
	prob = \frac{\#(corr\_class\_leaf\_examples)}{\#(corr\_class\_leaf\_examples+incorr\_class\_leaf\_examples)}
	\label{eq:confidence}
\end{equation}


One of the most used decision tree learning algorithms is the C4.5 algorithm~\cite{Quinlan:1993}.
C4.5 relies on the normalized information gain
to choose, for each node of the tree, the feature to be used for splitting the set of examples. The feature with the highest normalized information gain is chosen to make the decision.


\section{Approach}
\label{sec:approach}

In this section, we present the details of the proposed approach, which combines different existing techniques ranging from clustering approaches to decision tree learning, to provide predictions, at runtime, about the achievement of business goals in an execution trace.
In the following sections, we provide an overview of the approach and of the more specific implementation.



\subsection{General Approach}
\label{ssec:gen_approach}

Before presenting the approach proposed in this paper, some assumptions should be made. First, we assume that a set of historical execution traces of the process is available from which we can extract information about how the process was executed in the past. Based on the information extracted from the historical traces, we can provide predictions and recommendations for a running execution trace. Second, we assume that the underlying business process should be in some way non-deterministic or, at least, the mechanisms that guide the decisions taken during the process execution should not be known by the user. Any recommendation or prediction would be useless if the process participant already knows how the process develops given the input data values provided (we can think to a doctor who may not know about new therapies, or to a company providing services that does not know about the behaviors of its customers). Third, we assume that data used in the process are globally visible throughout the whole process.

\begin{figure}[t!]
	\begin{center}
		\includegraphics[width=\textwidth]{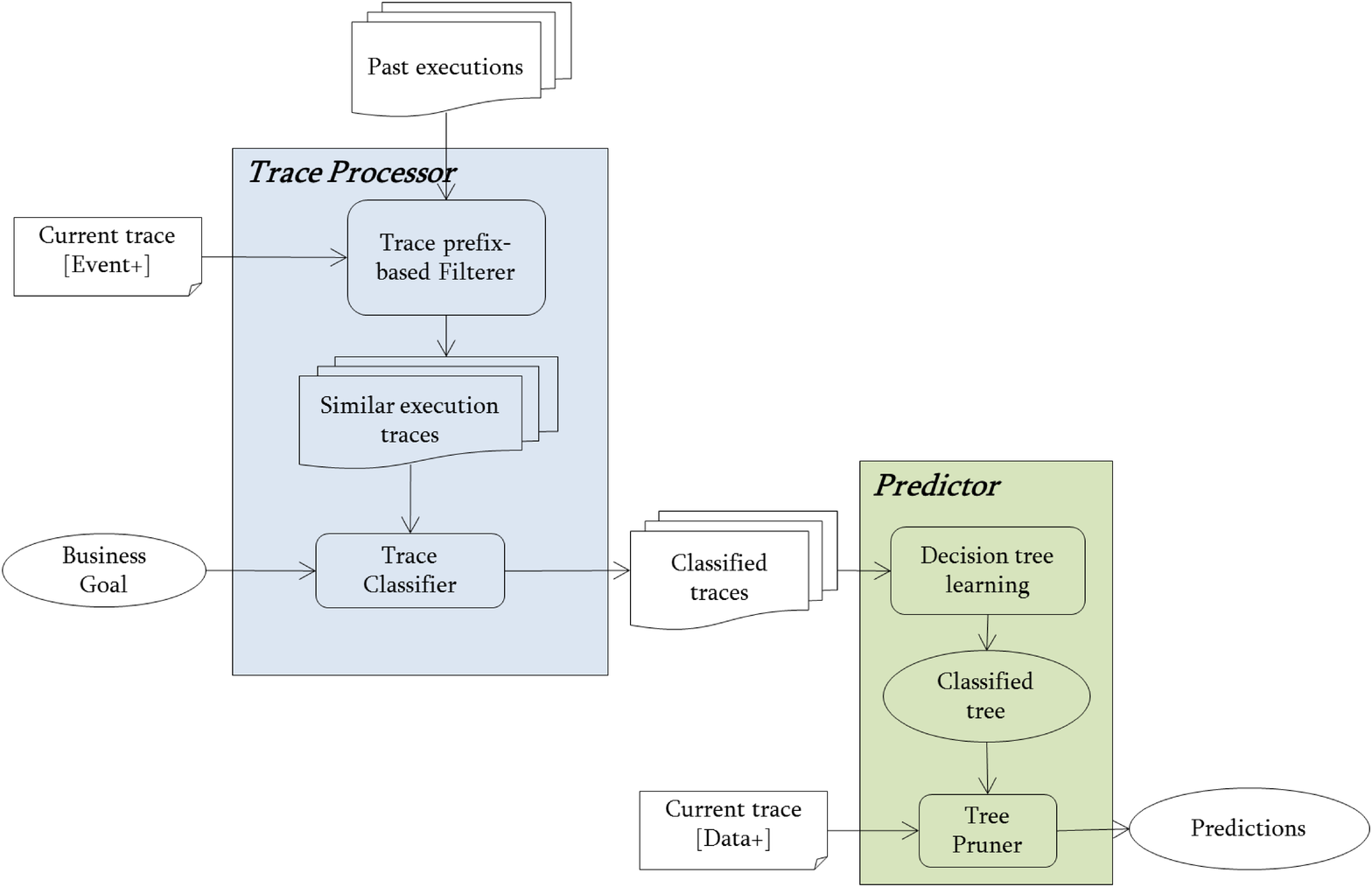}
	\end{center}
	\caption{\framework: architectural overview.}
	\label{fig:architecture}
\end{figure}

\figurename~\ref{fig:architecture} sketches the proposed \framework. It relies on two main modules: a \textit{Trace Processor} module to filter and classify (past) execution traces and a \textit{Predictor} module, which uses the \textit{Trace Processor} output as training data to provide predictions and recommendations (when an input is requested to the user).

The \textit{Trace prefix-based Filterer} submodule of the \textit{Trace Processor} module extracts from the set of historical traces only those traces having a prefix control flow similar to the one of the current execution trace (up to the current event). The filtering is needed since data values are usually strongly dependent on the control flow path followed by the specific execution. In addition, traces with similar prefixes are more likely to have, eventually in the future, a similar behavior. The similarity between two traces is evaluated based on their edit distance. We use this abstraction (instead of considering traces with a prefix that perfectly matches the current partial trace) to guarantee a sufficient number of examples to be used for the decision tree learning. In particular, a \emph{similarity threshold} can be specified to include more traces in the training set (by considering also the ones that are less similar to the current trace).

Each (historical) trace is identified with a \emph{data snapshot} containing the assignment of values for each attribute in the corresponding selected prefix.
The traces (the data snapshots) of the training set are classified by the \textit{Trace Classifier} submodule based on whether, in each of them, the desired business goal is satisfied or not. The goal is expressed in terms of a set of LTL formulas. 
In the case of our running example, the goal ``whenever a diagnosis is performed, then the patient will eventually recover'' can be represented in LTL through formula $\varphi_0$ reported in Section~\ref{sec:ltl}.


Formulas have to be satisfied along the whole execution trace. Four possible cases can occur at evaluation time:
\begin{itemize}
\item the formula is permanently violated: the prediction is trivial (non-satisfied);
\item the formula is permanently satisfied: the prediction is trivial (satisfied);
\item the formula is temporary violated/satisfied: the prediction should be able to indicate whether the formula will be satisfied or not in the future.
\end{itemize}

\begin{figure}[t!]
	\begin{center}
		\includegraphics[width=\textwidth]{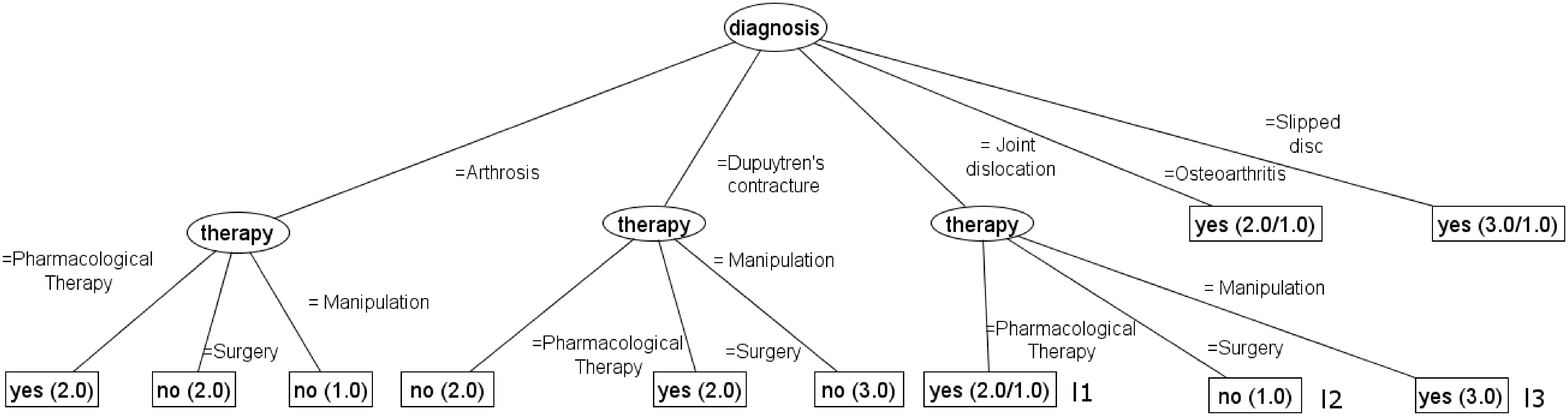}
	\end{center}
	\caption{Example decision tree.}
	\label{fig:tree}
\end{figure}

Once the relevant traces and, therefore, the corresponding data snapshots, are classified, they are passed to the \textit{Decision tree learning} module, in charge to derive the learned decision tree with the associated class support and probability. \figurename~\ref{fig:tree} shows a decision tree related to our running example: the number of data training examples (with values of the input variables following the path from the root to each leaf) respectively correctly and non-correctly classified is reported on the corresponding leaf of the tree. For example, for values ``Joint dislocation'' and ``Pharmacological therapy'', the resulting class is the formula satisfaction (``yes''), with 2 examples of the training set following the same path correctly classified and 1 non-correctly classified, i.e., with a class probability $prob = \frac{2}{2+1} = 0.66$.

All the data values assigned in the past, are supposed to be known by the predictor system at the current execution point of the trace. The tree can hence be pruned by removing all the branches corresponding to known values. The pruning algorithm returns either a unique path (and a unique class) or a subtree of the original tree, according to whether the system is used as predictor (the values of all the tree attributes are known) or as a recommender (there are attributes in the tree that are still unknown), respectively. In the latter case, leaves are ranked according to the associated class probability. The conditions on the values of the unknown attributes corresponding to the leaves with the highest rankings are returned to the user as recommendations.

For example, consider the case in which a diagnosis (``Joint dislocation'') and a therapy (``Pharmacological therapy'') have been given by the doctor. The \textit{Predictor} will consider only the path from the root to leaf $l_1$ (pruning all the other branches) and will predict the satisfaction of the formula with a probability class $prob = 0.66$ (see \figurename~\ref{fig:tree}).
We can also consider the case in which a diagnosis has already been made (e.g., ``Joint dislocation''), but no therapy has been prescribed yet. Then, all the branches corresponding to other values of the diagnosis attribute (i.e., ``Arthrosis'', ``Dupuytren's contracture'', ``Osteoarthritis'', ``Slipped disc'') can be pruned. Only the subtree corresponding to the branch ``Joint dislocation'' is analyzed and, since no other attribute is known, the class probability of each leaf computed. As shown in \figurename~\ref{fig:tree}, the three leaves have the following classes and class probabilities:
\begin{itemize}
\item $l_1$: satisfied with $prob_{l1} = \frac{2}{2+1} = 0.66$
\item $l_2$: non-satisfied with $prob_{l2} = \frac{1}{1} = 1$
\item $l_3$: satisfied with $prob_{l3} = \frac{3}{3+0} = 1$
\end{itemize}
The system will hence recommend ``Manipulation'' ($prob_{l3} = 1$).

Note that, if we consider as a feature of the decision tree the next activity to be executed, our framework is also able to recommend which activity should be performed next to maximize the probability of achieving a business goal.

\subsection{Implementation}

The approach has been implemented in the ProM process mining toolset. ProM provides a generic Operational Support (OS) environment~\cite{DBLP:conf/caise/AalstPS10,Westergaard2011:OS} that allows the tool to interact with external workflow management systems at runtime. A stream of events coming from a workflow management system is received by an OS service. The OS service is connected to a set of OS providers implementing different types of analysis that can be performed online on the stream. Our \textit{\framework} has been implemented as an OS provider.

\begin{figure}[t!]
	\begin{center}
		\includegraphics[width=0.65\textwidth]{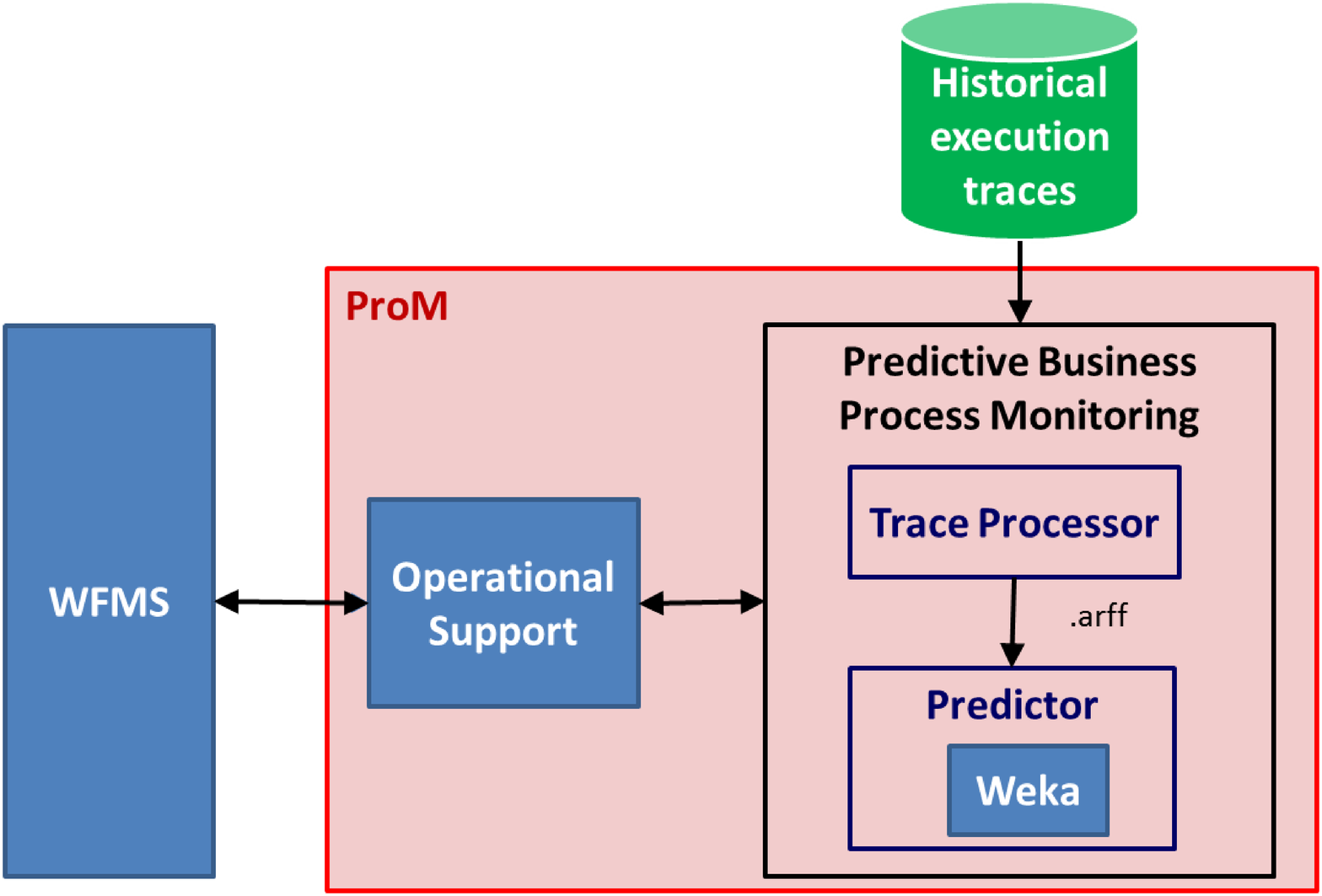}
	\end{center}
	\caption{\framework: implemented architecture.}
	\label{fig:impl}
\end{figure}

\figurename~\ref{fig:impl} shows the entire architecture. The OS service receives a stream of events (including the current execution trace) from a workflow management system and forwards it to the \textit{\framework} that returns back predictions and recommendations. The OS service sends these results back to the workflow management system.

For the implementation of the \textit{Predictor}, we rely on the WeKa J48 implementation of the C4.5 algorithm, which takes as input a .arff file and builds a decision tree. The .arff file contains a list of typed variables (including the target variable) and, for each trace prefix (i.e., for each data snapshot), the corresponding values. This file is created by the \textit{Trace Processor} and passed to the \textit{Predictor}.
The resulting decision tree is then analyzed to generate predictions and recommendations.

\section{Experimentation}
\label{sec:evaluation}

We have conducted a set of experiments by using the BPI challenge 2011 \cite{bpichallenge2011} event log. This log pertains to a healthcare process and, in particular, contains the executions of a process related to the treatment of patients diagnosed with cancer in a large Dutch academic hospital.
The whole event log contains $1,143$ cases and $150,291$ events distributed across $623$ event classes (activities). Each case refers to the treatment of a different patient. The event log contains domain specific attributes that are both case attributes and event attributes in addition to the standard XES attributes.\footnote{XES (eXtensible Event Stream) is an XML-based standard for event logs proposed by the IEEE Task Force on Process Mining (\url{www.xes-standard.org}).} For example, \emph{Age}, \emph{Diagnosis}, and \emph{Treatment code} are case attributes and \emph{Activity code}, \emph{Number of executions}, \emph{Specialism code}, and \emph{Group} are event attributes.

In our experimentation, first, we have ordered the traces in the log based on the time at which the first event of each trace has occurred. Then, we have splitted the log in two parts. We have used the first part (80\% of the traces) as training set, i.e., we have used these traces as historical data to derive predictions. We have implemented a log replayer to simulate the execution of the remaining traces (remaining 20\%) and send them as an event stream to the OS service in ProM (test set).

We defined 5 business goals corresponding to a subset (from $\varphi_1$ to $\varphi_5$) of the LTL rules reported in Section~\ref{sec:ltl}. This set of rules, indeed, allows us to exercise all the LTL constructs while investigating possibly real business goals.
We have asked for a prediction about each of the defined business goals in different evaluation points during the replay of each trace in our test set. In particular, we have considered as evaluation points the initial event (start event) of each trace, an early event (i.e., an event located at about 1/4 of each trace), and an intermediate event (i.e., an event located in the middle of each trace).

\begin{figure}[t!]
	\centering
		\includegraphics[width=1.\textwidth]{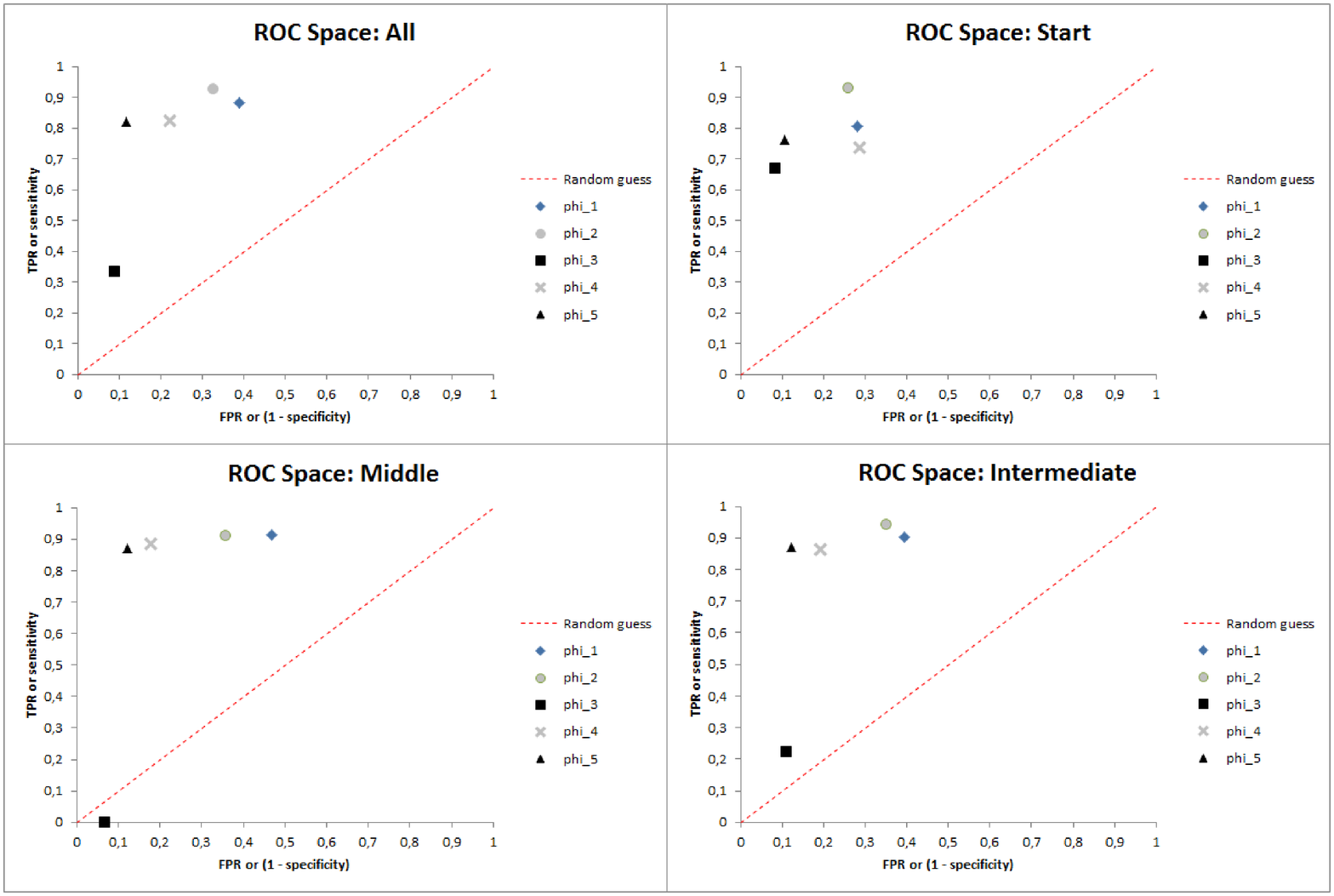}
	\caption{ROC spaces drawn for different LTL formulas and different evaluation points. Similarity threshold: 0.8; Minimum number of traces: 30.}
	\label{fig:ROC30}
\end{figure}

As well as a similarity threshold (see Section~\ref{ssec:gen_approach}), the implemented OS provider allows the user to specify a minimum number of traces to be used in the training set. In this way, if the threshold does not guarantee a sufficient number of examples, further traces are considered from the set of historical traces with a similarity with the current execution trace lower than the specified threshold.
In a first experiment, we have considered a similarity threshold of $0.8$ and a minimum number of traces of $30$.

For evaluating the effectiveness of our approach, we have used the ROC space analysis.
In particular, we have classified predictions in four categories, i.e., \begin{inparaenum}[\itshape i\upshape)]
\item true-positive ($T_P$: positive outcomes correctly predicted);
\item false-positive ($F_P$: negative outcomes predicted as positive);
\item true-negative ($T_N$: negative outcomes correctly predicted);
\item false-negative ($F_N$: positive outcomes predicted as negative).
\end{inparaenum}
The \emph{gold standard} used as reference is the set of all true positive
instances. In our experiments, we can easily identify the true positive instances. Indeed, if we are asking for a prediction at a certain point in time during the replay of a trace, we can understand if the prediction is correct by replaying the trace until the end.

To draw a ROC space, we need two metrics, i.e., the \emph{true positive rate (TPR)}, represented on the y axis, and the \emph{false positive rate (FPR)}, represented on the x axis. The TPR (or recall) defines how many positive outcomes are correctly predicted among all positive examples available:
\begin{equation}
	\textit{TPR} = \frac{T_P}{T_P + F_N}.
\end{equation}
On the other hand, the FPR defines how many negative outcomes are predicted as positive among all negative examples available:
\begin{equation}
	\textit{FPR} = \frac{F_P}{F_P + T_N}.
\end{equation}

\begin{table*}[t!]
	\caption{Evaluation of the approach for different LTL formulas, different evaluation points. Similarity threshold: 0.8; Minimum number of traces: 30.}
	\label{tab:ROC30}
	\centering
	\scalebox{0.75}{
	\begin{tabular}{c*{9}{>{\centering\arraybackslash}m{1.2cm}}}
    \toprule
    \specialcell{\textbf{}} &
    \specialcell{\textbf{TP}} &
    \specialcell{\textbf{FP}} &
    \specialcell{\textbf{FN}} &
    \specialcell{\textbf{TN}} &
    \specialcell{\textbf{TPR}} &
    \specialcell{\textbf{FPR}} &
    \specialcell{\textbf{PPV}} &
    \specialcell{\textbf{F1}} &
    \specialcell{\textbf{ACC}} \\
    \midrule
    \toprule
     \textbf{$\varphi_1$} &  &  &  &  &  &  &  &  &   \\
     \midrule
    \emph{Start} & 46 & 18 & 11 & 46 & 0.807 & 0.281 & 0.718 & 0.76 & \textbf{0.76}  \\
    \emph{Early} & 73 & 37 & 7 & 42 & 0.912 & 0.468 & 0.663 & 0.768 & \textbf{0.723}  \\
    \emph{Intermediate} & 75 & 34 & 8 & 52 & 0.903 & 0.395 & 0.688 & 0.781 & \textbf{0.751}  \\
    \emph{All} & 194 & 89 & 26 & 140 & 0.881 & 0.388 & 0.685 & 0.771 & \textbf{0.743}  \\
    \toprule
 \textbf{$\varphi_2$} &  &  &  &  &  &  &  &  &   \\
  \midrule
    \emph{Start} & 104 & 12 & 8 & 34 & 0.928 & 0.26 & 0.896 & 0.912 & \textbf{0.873}  \\
    \emph{Early} & 101 & 19 & 10 & 34 & 0.909 & 0.358 & 0.841 & 0.874 & \textbf{0.823}  \\
    \emph{Intermediate} & 110 & 19 & 7 & 35 & 0.94 & 0.351 & 0.852 & 0.894 & \textbf{0.847}  \\
    \emph{All} & 315 & 50 & 25 & 103 & 0.926 & 0.326 & 0.863 & 0.893 & \textbf{0.847}  \\
    \toprule
     \textbf{$\varphi_3$} &  &  &  &  &  &  &  &  &   \\
      \midrule
     \emph{Start} & 8 & 13 & 4 & 140 & 0.666 & 0.084 & 0.38 & 0.484 &\textbf{0.896}  \\
    \emph{Early} & 0 & 11 & 9 & 148 & 0 & 0.06 & 0 & 0 &\textbf{0.88}  \\
    \emph{Intermediate} & 2 & 18 & 7 & 143 & 0.222 & 0.111 & 0.1 & 0.137 & \textbf{0.852}  \\
    \emph{All} & 10 & 42 & 20 & 431 & 0.333 & 0.088 & 0.192 & 0.243 & \textbf{0.876} \\
    \toprule
     \textbf{$\varphi_4$} &  &  &  &  &  &  &  &  &   \\
      \midrule
     \emph{Start}& 53 & 33 & 19 & 82 & 0.736 & 0.286 & 0.616 & 0.67 & \textbf{0.721}  \\
     \emph{Early} & 54 & 18 & 7 & 83 & 0.885 & 0.178 & 0.75 & 0.812 & \textbf{0.845}  \\
     \emph{Intermediate} & 57 & 22 & 9 & 92 & 0.863 & 0.192 & 0.721 & 0.786 & \textbf{0.827}  \\
     \emph{All} & 164 & 73 & 35 & 257 & 0.824 & 0.221 & 0.691 & 0.752 & \textbf{0.795}  \\
 \toprule
     \textbf{$\varphi_5$} &  &  &  &  &  &  &  &  &   \\
      \midrule
     \emph{Start}& 55 & 10 & 17 & 85 & 0.763 & 0.105 & 0.846 & 0.802 & \textbf{0.838}  \\
     \emph{Early} & 52 & 13 & 11 & 94 & 0.825 & 0.121 & 0.8 & 0.812 & \textbf{0.858}  \\
     \emph{Intermediate} & 61 & 14 & 9 & 100 & 0.871 & 0.122 & 0.813 & 0.841 & \textbf{0.875}  \\
     \emph{All} & 168 & 37 & 37 & 279 & 0.819 & 0.117 & 0.819 & 0.819 & \textbf{0.857}  \\
    \bottomrule
	\end{tabular}}
\end{table*}

\begin{figure}[t!]
	\centering
		\includegraphics[width=1.\textwidth]{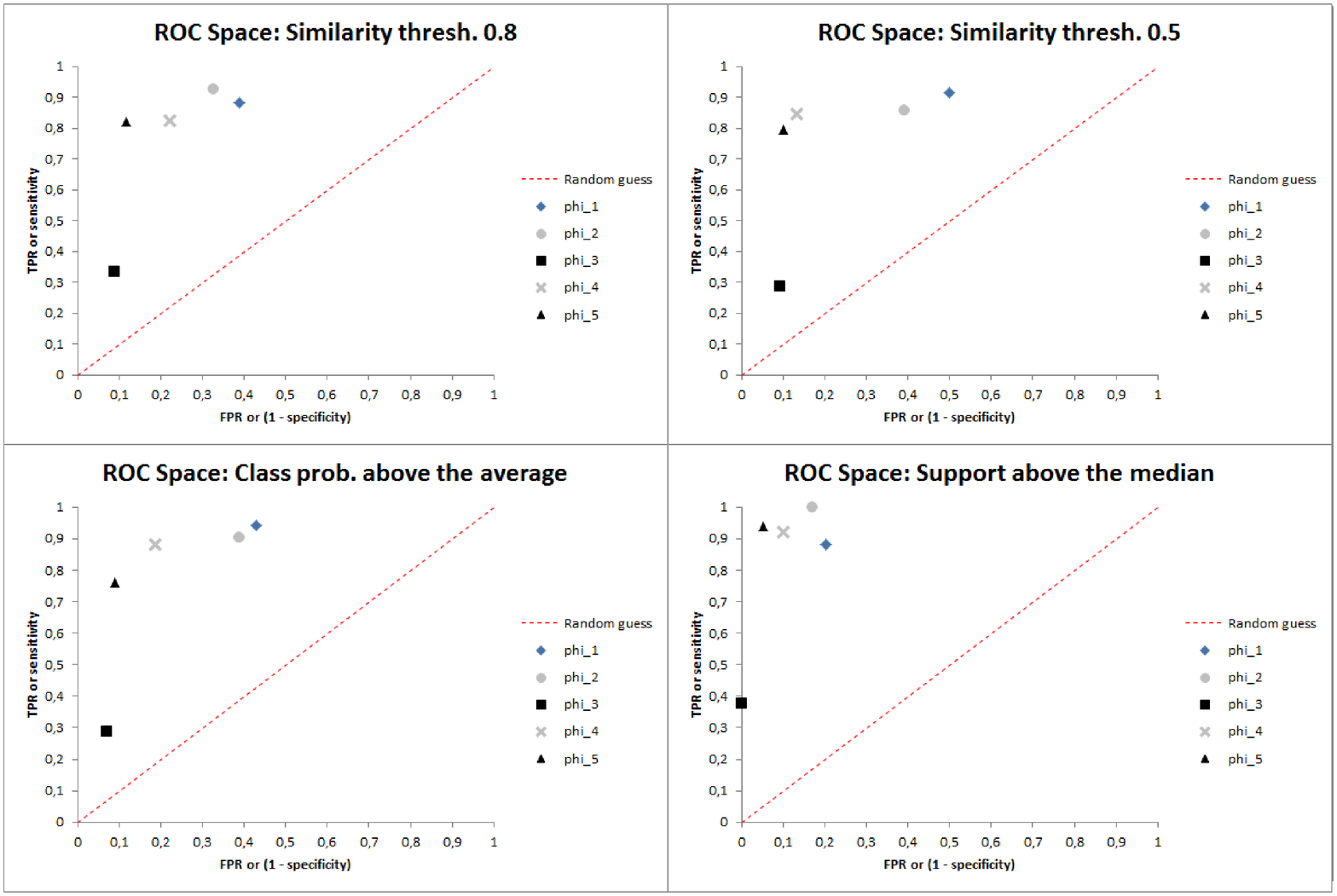}
	\caption{Comparison of ROC spaces drawn using (1) a similarity threshold of 0.8, (2) a similarity threshold of 0.5, (3) class probability higher than the average, and (4) class support higher than the median of the class supports.}
	\label{fig:ROC90}
\end{figure}

We have classified predictions for each LTL rule $\varphi_i$, and, therefore, each of them is represented as one point in the ROC space.
In \figurename~\ref{fig:ROC30}, we show four spaces drawn by classifying the evaluation points by position (start, early, intermediate). In the figure, we also show the results obtained by considering all the evaluation points together. Note that the best possible prediction method would yield a point in the upper left corner of the ROC space, representing $100\%$ sensitivity (no false negatives) and $100\%$ specificity (no false positives). A completely random guess would give a point along a diagonal line from the left bottom to the top right corners. Points above the diagonal represent good classification results, points below the line poor results.

The ROC space analysis highlights that for $\varphi_1$, $\varphi_2$, $\varphi_4$, and $\varphi_5$ our OS provider was able to discriminate well between positive and negative outcomes.\footnote{Note that, in some cases, the OS provider does not return any prediction. This is due to the fact that, when one of the features reported in the decision tree is an enumeration (and this is the case for several attributes in the considered log), it can happen that not all the possible values of the feature are included in a path from the root to a leaf of the decision tree. Therefore, it is not possible to do any prediction about the behavior of executions in which the feature has one of these values.} The results for $\varphi_3$ are less good since the number of positive examples for this formula is extremely low and the discovered decision tree overfits.

In general, the position in a trace in which we ask for a prediction does not affect significantly its reliability. In the presented scenario, in which case attributes are available since before the initial event occurs, this is true also for the initial event. Nevertheless, in case of overfitting, there is more variability. \tablename~\ref{tab:ROC30} shows that our results are good also in terms of \emph{positive predictive value (PPV)}, or precision, indicating how many positive outcomes are correctly predicted among all the outcomes predicted as positive:
\begin{equation}
	\textit{PPV} = \frac{T_P}{T_P + F_P},
\end{equation}
in terms of \emph{harmonic mean} of precision and recall:
\begin{equation}
F_1 =  2 \cdot \frac{\textit{PPV} \cdot \textit{TPR}}{\textit{PPV} + \textit{TPR}},
\end{equation}
and in terms of accuracy. Accuracy is particularly important in our context since it indicates how many times a prediction was correct:
\begin{equation}
	\textit{ACC} = \frac{T_P+T_N}{T_P + F_P + T_N + F_N}
\end{equation}
Note that the accuracy value is good also in case of overfitting (formula $\varphi_3$).

In a second experiment, we used a lower similarity threshold ($0.5$) and, again, a minimum number of traces equal to $30$. The results for this experiment (for all the evaluation points together) are reported in \tablename~\ref{tab:ROC90} and in \figurename~\ref{fig:ROC90}. This experiment shows that generating predictions based on a higher number of historical traces not always improves the quality of the results. This is due to the fact that, even if we are considering a larger training set, this set also includes traces that are quite dissimilar from the current trace, thus producing misleading results.


\begin{table*}[tb!]
	\caption{Evaluation of the approach using (1) a similarity threshold of 0.8, (2) a similarity threshold of 0.5, (3) class probability higher than the average, and (4) class support higher than the median of the class supports.}
	\label{tab:ROC90}
	\centering
	\scalebox{0.75}{
	\begin{tabular}{c*{6}{>{\centering\arraybackslash}m{1.2cm}}}
    \toprule
    \specialcell{\textbf{}} &
    \specialcell{\textbf{TPR}} &
    \specialcell{\textbf{FPR}} &
    \specialcell{\textbf{PPV}} &
    \specialcell{\textbf{F1}} &
    \specialcell{\textbf{ACC}} &
    \specialcell{\textbf{LOSS}} \\
    \midrule
    \toprule
     \textbf{$\varphi_1$} &  &  &  &  &  &  \\
     \midrule
    \emph{Similarity thresh. 0.8} & 0.881 & 0.388 & 0.685 & 0.771 & \textbf{0.743} & - \\
    \emph{Similarity thresh. 0.5} & 0.915 & 0.498 & 0.612 & 0.734 & \textbf{0.693} & - \\
    \emph{Class prob. above the average} & 0.94 & 0.429 & 0.714 & 0.812 & \textbf{0.767} & 0.223 \\
    \emph{Support above the median} & 0.88 & 0.201 & 0.83 & 0.854 & \textbf{0.841} & 0.508 \\
    \toprule
 \textbf{$\varphi_2$} &  &  &  &  &  &   \\
  \midrule
\emph{Similarity thresh. 0.8} & 0.926 & 0.326 & 0.863 & 0.893 & \textbf{0.847} & - \\
    \emph{Similarity thresh. 0.5} & 0.858 & 0.391 & 0.831 & 0.844 & \textbf{0.781} & - \\
    \emph{Class prob. above the average} & 0.903 & 0.39 & 0.851 & 0.876 & \textbf{0.818} & 0.294  \\
    \emph{Support above the median} & 1 & 0.171 & 0.971 & 0.985 & \textbf{0.974} & 0.519 \\
    \toprule
     \textbf{$\varphi_3$} &  &  &  &  &  &  \\
      \midrule
\emph{Similarity thresh. 0.8} & 0.333 & 0.088 & 0.192 & 0.243 & \textbf{0.876} & - \\
    \emph{Similarity thresh. 0.5} & 0.285 & 0.092 & 0.188 & 0.227 & \textbf{0.864} & - \\
    \emph{Class prob. above the average} & 0.285 & 0.07 & 0.176 & 0.218 & \textbf{0.897} & 0.167  \\
    \emph{Support above the median} & 0.375 & 0 & 1 & 0.545 & \textbf{0.977} & 0.559 \\
    \toprule
     \textbf{$\varphi_4$} &  &  &  &  &  & \\
      \midrule
    \emph{Similarity thresh. 0.8} & 0.824 & 0.221 & 0.691 & 0.752 & \textbf{0.795} & - \\
    \emph{Similarity thresh. 0.5} & 0.846 & 0.132 & 0.754 & 0.797 & \textbf{0.86} & - \\
    \emph{Class prob. above the average} & 0.881 & 0.186 & 0.728 & 0.797 & \textbf{0.838} & 0.206  \\
    \emph{Support above the median} & 0.92 & 0.1 & 0.793 & 0.851 & \textbf{0.905} & 0.518  \\
      \toprule
     \textbf{$\varphi_5$} &  &  &  &  &  & \\
      \midrule
\emph{Similarity thresh. 0.8} & 0.819 & 0.117 & 0.819 & 0.819 & \textbf{0.857} & - \\
    \emph{Similarity thresh. 0.5} & 0.794 & 0.101 & 0.807 & 0.801 & \textbf{0.862} & - \\
    \emph{Class prob. above the average} & 0.761 & 0.089 & 0.809 & 0.784 & \textbf{0.86} & 0.23  \\
    \emph{Support above the median} & 0.938 & 0.053 & 0.938 & 0.938 & \textbf{0.942} & 0.53 \\
    \bottomrule
	\end{tabular}}
\end{table*}

One way of assessing the reliability or ``goodness'' of a prediction is to use its \emph{class probability}. In \tablename~\ref{tab:ROC90} and in \figurename~\ref{fig:ROC90}, we show the results obtained by filtering out predictions with a class probability that is lower than the average. \tablename~\ref{tab:ROC90} also reports the prediction loss (LOSS), i.e., the percentage of predictions lost when filtering out predictions with a low class probability. This experiment shows that considering only predictions with a high class probability not always improves the quality of the results, though the percentage of predictions lost is not high (about 20\%).

Another way of evaluating the reliability of a prediction is to consider its class support. In \tablename~\ref{tab:ROC90} and in \figurename~\ref{fig:ROC90}, we show the results obtained by filtering out predictions with support lower than the median of the supports. In this case, although the cut of predictions is
high (more than half of the predictions are filtered out), there is a clear improvement in all the considered metrics:
in the ROC dimensions, in the F-measure as well as in the average accuracy of the predictions.





In summary, the evaluation shows that the proposed approach is feasible and provides accurate predictions (and hence recommendations).
Results seem overall not to be affected by the position of the evaluation point, thus demonstrating that the approach works well even when few variables are known.
Support seems to be an important factor influencing the results, i.e., the more evidences we have in the training set, the more accurate are the produced predictions. If on the one hand this highlights the need to have adequate training sets, on the other it also shows that sacrificing outlier predictions, it is possible to obtain very accurate results (accuracy around 0.9).

\section{Related Work}
\label{sec:related}
In the literature, there are some works that provide approaches for generating predictions and recommendations during process execution and are focused on the time perspective. In \cite{DBLP:journals/is/AalstSS11,DBLP:conf/caise/AalstPS10}, the authors present a set of approaches based on annotated transition systems containing time information extracted from event logs. The annotated transition systems are used to check time conformance while cases are being executed, predict the remaining processing time of incomplete cases, and recommend appropriate activities to end users working on these cases. In \cite{Folino}, an ad-hoc predictive clustering approach is presented, in which context-related execution scenarios are discovered and modeled through state-aware performance predictors.

There are several works focusing on generating predictions and recommendations to reduce risks. For example, in \cite{DBLP:conf/caise/ConfortiLRA13}, the authors present a technique to support process participants in making risk-informed decisions, with the aim of reducing the process risks. Risks are predicted by traversing
decision trees generated from the logs of past process executions. In \cite{Pika}, the authors propose an approach for predicting of time-related process risks by identifying (using statistical principles) indicators observable in event logs that highlight the possibility of transgressing deadlines. In \cite{suriadi}, the authors propose an approach for Root Cause Analysis based on classification algorithms. After enriching a log with information like workload, occurrence of delay and involvement of resources, they use decision trees to identify the causes of overtime faults.


A key difference between these approaches and our technique is that they rely either on the control-flow or on the data perspective for making predictions at runtime, whereas we take both perspectives into consideration. In addition, the purpose of our recommendations is different. We provide recommendations neither to reduce risks nor to satisfy/discover timing constraints. We aim instead at maximizing the likelihood of achieving business goals expressed in the form of LTL rules.


\section{Conclusion}
\label{sec:conclusion}

This paper presented a framework for predictive business process monitoring based on the estimation of probabilities of fulfillment of LTL rules at different points during the execution of a case. The framework takes into account both the sequencing of activities as well as data associated to the execution of each activity. A validation of the framework using a real-life log demonstrates that recommendations generated based on the framework have a promising level of accuracy when sufficient support is available.

Increased accuracy could be achieved by extending the technique along two directions. First, the proposed technique matches the trace of an ongoing case against prefixes of completed traces based on edit distance. While this is a well-known measure of similarity and suitable as a first step in this study, other approaches could be considered, including trace similarity measures based on occurrences of n-grams, counts of activities and activity pairs, and other relevant features that have been studied in the context of trace clustering~\cite{MedeirosGGAWDS07}. In a similar vein, discriminative sequence mining techniques~\cite{Lo2011} could be applied in order to extract prefix patterns that are associated with fulfillment of a given business goal. These patterns can also be taken as input in the prediction.
Secondly, we have considered the use of decision trees to build the classifier. With larger number of attributes, which might be encountered in richer logs, decision trees are likely to exhibit lower accuracy due to their inherent weaknesses when dealing with large feature sets.
In this context, other classification techniques, such as random forests or sparse logistic regression are possible alternatives.

\smallskip\noindent
\textbf{Acknowledgments.} This work is partly funded by ERDF via the Estonian Centre of Excellence in Computer Science.

\bibliographystyle{splncs03}
\bibliography{CAISE2014}

\end{document}